\documentstyle[prl,aps,psfig,epsfig,multicol]{revtex}
\begin{document}
\draft
\title{Superconductivity in Na$_x$CoO$_2$$\cdot$yH$_2$O :\\
Is p-wave pairing less likely because of a competing charge order ?}
\title{Superconductivity in Na$_x$CoO$_2$$\cdot$yH$_2$O : 
\\Is Spin-Charge Separation Protecting a d$_1$+id$_2$ State ?}

\author{Debanand Sa$^1$, Manas Sardar$^2$ and G. Baskaran$^3$,}
\address{ $^{1,3}$The Institute of Mathematical Sciences, C. I. T. Campus,
Chennai-600 113, India \\
$^2$Materials Science Division, Indira Gandhi Centre for
Atomic Research, Kalpakkam-603 102, India }


\maketitle

\begin{abstract}
Superconductivity in Na$_x$CoO$_2$$\cdot$yH$_2$O is likely to be a 
p or d-wave; however, experiments are unable to pinpoint the symmetry. 
A simple estimate of pair breaking effects from an unavoidable `Na$^+$ 
vacancy disorder' in an ordered Na$^+$ lattice, at an 
optimal $x_{\rm opt} \approx 0.30$
is shown to destroy a  
Fermi liquid based p or d-wave superconductivity.  However, a robustness 
of superconducting and normal states, seen in experiments is pointed out 
and argued to imply presence of a `quantum protectorate', possibly a 
`spin-charge decoupling' that protects a d$_1$+id$_2$ and not a p-state.  
A calculation of Knight shift and $1 \over T_1$ in the framework of 
RVB mean field theory and a fit to the data of Kobayashi 
${\it et \> al.}$ \cite{kobayashi103} is made.
\end{abstract}

\begin{multicols}{2}
\narrowtext

The recent discovery of superconductivity (SC) in
Na$_x$CoO$_2$$\cdot$yH$_2$O
\cite{takada03} has generated renewed interest, due to the role of
strong electronic correlations, in the cobalt oxide family. This
compound has layered  crystal structure where Co atoms form a two
dimensional triangular lattice separated by Na$^{+}$ ions and H$_2$O
molecules. A number of experimental
results \cite{terasaki97,sakurai03,cao03,jin03,chou03,schaak03,yayu03,kobayashi103,waki03,fujimoto03,ishida03,lynn03,jorgensen03} 
in these family of material 
are quite anomalous and remains unexplained.

A problem of current interest is the symmetry of the superconducting
order parameter. In known examples, where electron correlation 
provides a superconductivity mechanism,
the order parameter symmetry is unconventional, either d or p-wave.
Nuclear spin resonance experiments, which can identify order parameter
symmetry, has so far not been decisive on this
issue for Na$_x$CoO$_2$$\cdot$yH$_2$O. On the theoretical side, a single 
band t-J
model introduced by one of us \cite{baskaran103} allows, at small dopings,
a spin-charge decoupled normal state and a resonating valence bond (RVB) 
superconductivity with d$_1$+id$_2$ symmetry; and a triplet pairing 
at high dopings, arising from a ring exchange processes and a consequent
ferromagnetism induced by dopant dynamics. Subsequent 
papers\cite{brijesh03,wang03,ogata03} 
have focussed on RVB theory of d-wave pairing and 
others\cite{singh00,tanaka03,ikeda03} on triplet
pairing invoking certain ferromagnetic tendency.

The one band t-J model physics introduced in reference\cite{baskaran103} 
implies a distinct possibility of `spin-charge decoupling' for a wide
range of doping. An important consequence, as we have learned from our 
study of cuprates, is a remarkable possibility of a 
`quantum protectorate' \cite{anderson00,laughlin00}. 
That is, an ability to protect superconducting and non-Fermi liquid normal 
states from {\em phonons and disorder}. This issue needs to be investigated 
in Na$_x$CoO$_2$$\cdot$yH$_2$O.
However, a quantum protectorate aspect is visible in some experiments:
i) replacement of H$_2$O by D$_2$O, that alters the phonon frequency of 
ice layers substantially, in the experiments of Jin ${\it et \> al.}$ 
\cite{jin03} 
has no appreciable effect either on normal state resistivity or the 
superconducting T$_c$ ii) survival of superconductivity in a narrow 
dome ($ {1\over 4} < x < {1\over 3})$ \cite{schaak03}, overcoming a
possible strong competition from charge 
order \cite{baskaran203}, a charge localization tendency,
iii) unavoidable disorder in the Na-H$_2$O layers \cite{lynn03,jorgensen03}, 
another charge localization tendency. A large ab-plane residual resistivity,  
$\rho_{ab}\sim$ 300 $\mu\Omega$ cm is observed in a first single crystal 
(T$_c$ $\sim$ 5 K) measurement \cite{jin03}; 
even if a fraction of it were intrinsic, 
a Fermi liquid based d or p-wave would not have survived. For example, 
in Sr$_2$RuO$_4$, a well established p-wave superconductor 
\cite{mackenzie03} even a small disorder corresponding to a residual 
resistance of $\rho_{ab}\approx$ 1.5 $\mu\Omega$ cm is capable of 
suppressing a superconducting T$_c$ ($\approx$ 1.5 K) to zero.    

After summarizing known phenomenological evidence for p or d-wave
superconductivity, we argue for the presence of an 
{\em unavoidable Na$^+$ vacancy disorder} in an ordered  
Na$^+$ plane at optimal
doping $x = x_{\rm opt} \approx 0.30$. We quantify the disorder and 
show that it is strong enough to destroy a Fermi liquid based p or 
d-wave superconductivity. We then discuss how spin-charge decoupling is 
capable of protecting a d-wave and not a p-wave from the singular effects 
of disorder known from Anderson's theorem of dirty superconductors.  A 
simple RVB mean field theory results for Knight shift and $1\over T_1$
and comparison with experimental results of Kobayashi ${\it et \> al.}$  
\cite{kobayashi103} is     

\begin{table*}
\caption{Summary of the Knight shift and the nuclear spin relaxation
rate for Na$_x$CoO$_2$$\cdot$yH$_2$O:}
\label{cryst}
\begin{tabular}{|c|c|c|c|c|}
References   & $1\over T_1$ (T$_c$) & $K\%$ (T$_c$) & Coherence & Yoshida \\
          &                      &               &  peak     & function\\
\hline
Kobayashi &   61 sec$^{-1}$      &   3.1         &  Yes      &  Yes    \\
et al.[9]&                      &               &           &         \\
\hline
Waki      &   1 sec$^{-1}$       &   1.8         &  Yes      &   No    \\
et al.[10]&                      &               &           &         \\
\hline
Fujimoto  &   60 sec$^{-1}$      &               &   No      &         \\
et al.[11]&                      &               &           &         \\
\hline
Ishida    &   70 sec$^{-1}$      &               &   No      &         \\
et al.[12]&                      &               &           &         \\
\end{tabular} 
\end{table*} 

\noindent made. 

Various measurements have looked for unconventional superconductivity  
in Na$_x$CoO$_2$$\cdot$yH$_2$O:
uniform spin susceptibility $\chi_s$, NMR/NQR, 
specific heat and resistivity. A summary of nuclear spin resonance results
\cite{kobayashi103,waki03,fujimoto03,ishida03} are shown in table 1.
Some of the marked differences in NMR/NQR results are likely to arise from 
sample quality. Uniform spin susceptibility measurements from all the
groups \cite{sakurai03,cao03,jin03,chou03,kobayashi103,fujimoto03,ishida03} 
has the form \cite{sakurai03}: $\chi_s \approx
\chi_0(T) + {C\over (T+\Theta)}$. Here $\chi_0(T)$ is a nearly 
temperature independent but enhanced Pauli component (enhanced by a 
factor of 5-7 compared to free electron value); this component
also exhibits a `spin gap' like reduction by about 5-10$\%$ from
250 K to T$_c$. The Curie-Weiss component has a large variation
between groups: single crystal measurements by two 
groups \cite{jin03,chou03} show a small value of Curie constant 
C $\sim$ 0.05 $\times$ 10$^{-2}$ emu K/mole 
compared to $\sim$ 1.0 $\times$ 10$^{-2}$ emu K/mole given by other groups 
from powder samples \cite{sakurai03,cao03,ishida03}. It is likely, 
as suggested in reference \cite{sakurai03,kobayashi103} that 
the Curie-Weiss component may not be intrinsic at optimal doping
in Na$_x$CoO$_2$$\cdot$yH$_2$O.

Nuclear spin relaxation $1\over T_1 T$ scales with $\chi_s$ \cite{ishida03},
for a range of temperatures above T$_c$. However, a Curie-Weiss 
enhancement in $1 \over T_1T$ seen below about 100 K is not seen by 
all groups.  Notably, Kobayashi ${\it et \> al.}$ \cite{kobayashi103} 
observe a ${ 1\over T_1T} \approx$ 12.5 sec$^{-1}$ K$^{-1}$ in the 
range T$_c$ $<$ T $<$ 30 K and
no significant enhancement. This result also questions the intrinsic 
character of Curie-Weiss enhancement in $1 \over T_1 T$. 

Specific heat coefficient $\gamma$ shows, in all experiments 
\cite{sakurai03,cao03,jin03,chou03,ueland03,lorenz03,yang03}, an 
enhancement over the free electron value suggesting a mass enhancement
by a factor about 7. One of the single crystal measurements of $\gamma$
and $\chi_s$ \cite{chou03} estimate a Wilson ratio 
R=${{4\pi^2 k_B^2\chi_s}\over{3(g\mu_B)^2\gamma}}$ $\approx$ 1.53. 
Accordingly they interpret the system as a heavy metal.

Putting measurements for T $>$ T$_c$ together we may say that there is 
no particular preference between a p or a d-wave, even though many
experimental groups suspect a p-wave. 

The situation below T$_c$ is unclear, with conflicting results for
coherence peak and additional Yoshida-function like Knight shift.   
However, it is fair to point out that Kobayashi ${\it et \> al.}$'s result is 
intriguing and demands a further discussion; i) in addition to
absence of Curie-Weiss enhancement above T$_c$, alluded to earlier,
there is a clear coherence peak below T$_c$ in $1\over T_1$ and 
ii) one sees an 
additional Knight shift and a Yoshida function type behavior below 
T$_c$. Is it possible that Curie-Weiss contributions are extrinsic 
and Kobayashi ${\it et \> al.}$'s sample has less of it and 
therefore has managed
bring out a coherence peak and non-trivial Knight shift ? It is important 
to investigate this issue further. 

At the present stage it is difficult to conclude about possible nodal
character of the order parameter from the power 
law behavior of $1 \over T_1$ \cite{fujimoto03} 
and specific heat \cite{lorenz03,yang03} 
below T$_c$, until we exclude extrinsic effects.

Putting measurements for T $<$ T$_c$ together, we may say that perhaps
there is a support for a gaplful d$_1$+id$_2$ or p$_1$+ip$_2$ pairing, 
unless the coherence peak and additional Knight shift measurements of 
Kobayashi ${\it et \> al.}$ \cite{kobayashi103} are proved wrong.


Triplet superconductors such as, Sr$_2$RuO$_4$, UGe$_2$ etc. are known 
to be ultrasensitive to non-magnetic (scalar) disorder.
In what follows, we make an estimate of pairbreaking effects from
two sources of disorder in Na$_x$CoO$_2$$\cdot$yH$_2$O at optimal
doping, within the context of Fermi liquid theory. We find, contrary 
to the experimental observation, that existing pair breaking effects
should suppress superconducting T$_c$ completely. 

In the theory of dirty superconductors \cite{abrikosov}, the parameter
$\alpha \equiv {\hbar\over{\tau k_B T_{c0}}}$ plays an important role;
here $1\over\tau$ is the `pair breaking' rate and $\hbar \over \tau$ 
is the energy spread an eigen state (a Bloch state), close to the Fermi 
level, acquires on introduction of disorder in the system. When this 
energy spread is nearly equal to $k_BT_{c0}$, i.e., $\alpha \approx 1$ 
superconductivity disappears. This also roughly corresponds to
the electron mean free path $\ell$ at the Fermi level becoming nearly 
equal to the coherence length $\xi$.  

Sources of pair breaking are i) an irreducible Na$^+$ vacant disorder 
at `optimal doping', ii) low frequency charge order fluctuations  
and iii) dissipation from the low frequency polar modes of the H$_2$O
layers. We first discuss the effect of disorder in Na$^+$ layer.
As elaborated in \cite{baskaran203} it is convenient to define a reference 
`stoichiometric' compound Na$_{1\over3}$CoO$_2\cdot{4\over3}$H$_2$O, 
where Na atoms are ordered; in an ideal ordered structure {\em they 
fill one of the three sub lattices of a triangular lattice of preferred 
sites and have a $\sqrt 3 \times \sqrt 3$ order}. 
This is consistent with the large coherence length of $1000~\AA$ 
in the ordered Na$^+$ layer seen in the structural study \cite{jorgensen03}. 
Experimentally optimal T$_c$ $\approx$ 5  is reached at 
$x_{\rm opt}={1\over 2}({1\over 4} + {1\over 3}) \sim 0.3$; that is,
a removal of $\delta x  \equiv
({1\over 3} - x_{\rm opt}) \approx {1\over24}$ of Na atoms from the 
ordered Na$^+$ layer. Assuming that 
these vacancies are disordered, we get an `irreducible' vacancy 
disorder in an otherwise ordered lattice of Na$^+$ ions.  
A corresponding missing `dipole moments' exists in the 
H$_2$O layer as well. 

At optimal doping, a fraction $ 2 \times {1\over 24}$ of vacant Na$^+$ 
sites from the sandwitching Na$^+$ layers above and below provide a 
random potential to 
the delocalized carriers in the CoO$_2$ planes. A change in 
potential at the nearest Co site, arising from a removal of one 
Na$^+$ is $\approx {e^2\over{R \epsilon_{\rm total}}} 
\approx $ 80 meV; here $R \approx 5.3~\AA$ and $\epsilon_{\rm total}
= \epsilon_{\infty} + \epsilon_{H_2 O} \approx$ 30. In the above
we have taken into account a large dielectric screening from the 
hydrogen bonded H$_2$O sheet \cite{baskaran203}. 
The strength of the above scattering
potential 80 meV is the same order of magnitude as the bandwidth 
200 meV as estimated in reference \cite{chou03}. This makes the
mean free path of an electron in the above tight binding band of the 
order of the average vacancy distance 
${1\over \sqrt \delta x} \sim 4 $ (in units of Co-Co distance). 

The above estimate of the mean free path is less than 
$\xi$($\sim 100~\AA$),
the coherence length in our superconductor. Thus we get the result
that $ \alpha < \alpha_{\rm critical} \approx 1$, making a Fermi liquid 
based p or d-wave superconductivity unlikely. Since the Na$^+$ vacancy
disorder is strong enough to suppress a d or p-wave superconducting
T$_c$, we do not discuss our estimate of additional pair breaking effects 
from charge order fluctuations and low frequency polar fluctuations of 
ice layers.

The fact that the superconductor has survived severe pair breaking
effects of disorder discussed above and a robustness with respect to
replacement of H$_2$O by D$_2$O \cite{jin03} suggests to us a 
possible existence 
of a `quantum protectorate'. If our 2D t-J modeling is correct, a 
spin-charge decoupled state is a likely consequence and it will become
a natural quantum protectorate. Spin-charge decoupling, though suspected, 
has not been experimentally investigated so far in 
Na$_x$CoO$_2$$\cdot$yH$_2$O, a strongly correlated 2D metal. The issue 
of quantum protectorate is well known in cuprates 
\cite{anderson00}, another strongly 
correlated 2D metal, through the robustness of the non-Fermi liquid 
normal state and $d_{x^2-y^2}$-wave superconductivity against 
off-plane disorder and phonons. 

Below we give a physical picture of spin-charge separated state as 
a quantum protectorate and how a singlet d-wave and not a 
triplet p-wave will be protected. First we recall how non-magnetic 
disorder affects pairing in a Fermi liquid: scattering of individual 
${\bf k}$ states of electrons by scalar disorder interferes destructively 
the coherent scattering of (${\bf k},-{\bf k}$) pair over the Fermi surface, 
whenever it is trying to establish a pair amplitude that changes sign along 
the Fermi surface, $\sum_{\bf k} \Delta({\bf k}) = 0$, as in the case 
of d or p-wave. 

In a t-J model, the 
singlet stabilising collision between two electrons is a superexchange
process that happens between two singly occupied sites (neutral spins),  
in our case between two 
adjacent spin-half moments Co$^{4+}$'s. The charge 2e condensation, 
namely superconductivity in RVB theory is viewed as taking place in two
steps: i) pairing between neutral spins in the spin singlet channel
and then condensation of holon or doublons. The quantum protectorate
hypothesis essentially states that the effect of scalar disorder on
spinon/holon/doublon (collective `topological' excitations) 
is in some sense `averaged out', and remain unaffected for small disorder. 
To this
extent a local superexchange process that involves two neutral spins 
remains unaffected. Similiarly a holon/doublon condensation that 
involves a spinless charge also remain unaffected. From this point of 
view, as long as the pairing is in spin-singlet channel it is compatible 
with the robust superexchange process. Thus singlet pairing such as an 
extended-s or d-wave can be protected by spin-charge decoupling.

Next important question is how a p-wave may not be protected by
spin-charge decoupling. A possible p-wave pairing suggested in 
reference\cite{baskaran103}, goes beyond superexchange and invokes 
a ring exchange \cite{anderson01} 
involving a minimum of four sites and 5 electrons 
and a consequent ferromagnetic tendency given by an effective 
J$_{\rm eff}$ $\approx$ J - $x|t|< 0 $ (when the hopping matrix element 
t has a `wrong' sign). This dopant induced dynamics involves charge 
transport. To this extent it is not a triplet pairing between two 
neutral spins. It is rather
a pairing between two electrons in the triplet channel. Thus 
spin-charge decoupling is unable to protect a triplet state,
that involves fragile electrons rather than the robust neutral spins
or doublons. 

Having offered some phenomenological and microscopic support 
for d$_1$+id$_2$ pairing, we have performed a simple RVB
d$_1$+id$_2$ mean field theory calculation for $1\over T_1$ and Knight
shift and fit to the experimental results of
Kobayashi ${et \>al.}$ \cite{kobayashi103}.
We  point out that Bang ${\it et \> al.}$ \cite{bang03} have
calculated the effects of impurities on $1\over T_1$
in the d$_1$+id$_2$ phase. We refer the readers to
reference\cite{baskaran103,brijesh03,wang03,ogata03} 
for details of RVB mean field theory.
After a quantitative fit to Kobayashi
${\it et \> al.}$'s  data, we get a value of hyperfine contact coupling
to be 39.32$\times$ 10$^{-4}$cm$^{-1}$.  Using the same hyperfine
coupling, we compute the temperature dependent nuclear spin relaxation
rate and make quantitative comparison with the existing experimental data.

Nuclear spin (${\bf I}_0$) relaxes by coupling with the 
local conduction electron spin density (${\bf S}_0$) 
in a metal. The interaction Hamiltonian is given as,
$H_{\rm hyp}=A{\bf I}_0\cdot{\bf S}_0$, 
where, $A$ is the effective hyperfine coupling. 
The spin part of the Knight shift is,
$$K_s={{1}\over{2}}({{\mu_B}\over{\mu_N}})A
\chi_s(q\rightarrow 0,\omega=0), $$ 

\noindent where, $\chi_s(q,\omega)$ is the dynamic spin susceptibility.
The temperature dependent order parameter in a RVB SC is computed
self consistently and is incorporated in the susceptibility to
get the temperature dependent Knight shift. We fix the
chemical potential $\mu$=0.075 eV to get 30$\%$ electron doping.
We also take $J$=6.2 meV in order to get a T$_c$=4.5 K. Throughout
the computation, we have chosen $t_{\rm eff}$=0.03 eV and a cut-off
energy scale to be 500 K.  The ratio $2 \Delta(T=0)/k_B T_c$ comes
out to be approximately 3.5.

In Fig. 1 we show the fit of the computed temperature dependent
Knight shift to the data of
Kobayashi ${\it et \> al.}$ \cite{kobayashi103} from which we extracted
the value of the hyperfine coupling parameter,
$A$=39.32$\times$ 10$^{-4}$ cm$^{-1}$.

In the process of fitting the data we subtracted a fairly large 
temperature independent background ($K_{orb} \%$) of 2.55$\%$ from
$K_{tot}$$\%$, where $K_{tot}(T) = K_{orb} + K_s$(T). And
$K_s(T=T_c)-K_s(T=0)$ is nearly 0.6$\%$. The absolute value of 
$K_{tot}\%$ in this case is nearly 3.15$\%$.
The value of $A$ estimated  as above compares well with
high T$_c$ cuprates\cite{mila89} whereas $K_{tot}(T)$ and $K_{orb}$
do not. In cuprates,  $K_{orb}$ was estimated to be 0.23$\%$ and 
$K_s(T=T_c)-K_s(T=0)=0.37 \%$ which are much smaller compared to 
layered cobaltate.
 
\begin{figure}
\psfig{figure=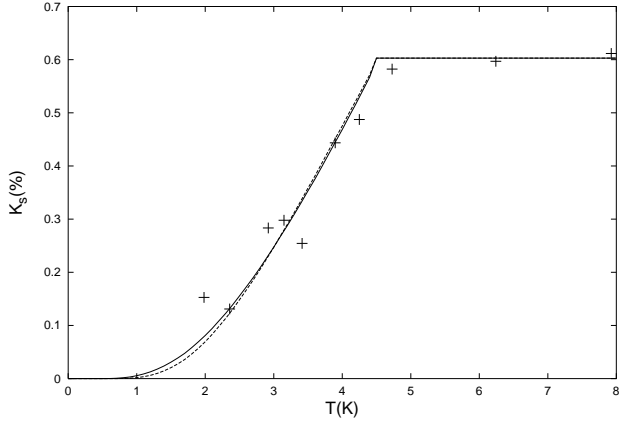,angle=-90,width=0.95\linewidth}
\caption{Fit of the computed Knight shift in a RVB SC 
with d$_1$+id$_2$ gap symmetry (solid lines) and
a BCS SC with s-wave gap (dashed lines) to the data
by Kobayashi ${\it et \> al}$. [9]. \label{knight}}
\end{figure}

\begin{figure}
\psfig{figure=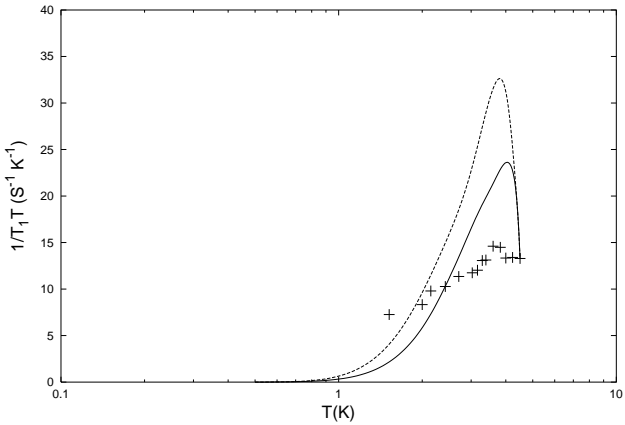,angle=-90,width=0.95\linewidth}
\caption{Computed nuclear spin relaxation in a RVB SC 
with d$_1$+id$_2$ gap symmetry (solid lines)(multiplied by 0.65) and
a BCS SC with s-wave gap (dashed lines) compared to the
data by Kobayashi ${\it et \> al}$. [9]. \label{nmrt}}
\end{figure}

The relaxation rate 
from the golden rule is given by,

$${1\over T_1}={{2 \pi}\over{\hbar}}{\mid A\mid}^2\sum_{k,q}l^2(k,q)
f_k(1-f_{k+q})\delta(E_{k+q}-E_k-\hbar\omega_0), $$   

\noindent where, the coherence factor $l^2$ in a d$_1$+id$_2$
superconductor is written as, $l^2(k,q)={{1}\over{2}}
(1+{{\epsilon_k\epsilon_{k+q}+\mid\Delta_k\mid\mid\Delta_{k+q}\mid}
\over{E_k E_{k+q}}})$. Here,  $\omega_0$ is the nuclear Zeeman frequency
which is extremely small on the scale of energy of interest.
In computing the relaxation time for a RVB superconductor, we
consider the same parameters as we used for computing the
Knight shift, also the same temperature dependence of the SC
order parameter. Further, the hyperfine coupling which we estimated
by fitting our data with the Knight shift measurement has been used to
compute the relaxation time.

Fig. 2 shows that the overall
shape of the spectrum looks similar to  the data of Kobayashi
${\it et \> al.}$ \cite{kobayashi103} and
Waki ${\it et \> al.}$ \cite{waki03} but with some differences.
We find that the value of ${{1}\over{T_1T}}$ at T$_c$ by  
Kobayashi ${\it et \> al.}$ \cite{kobayashi103} is about 0.65 times 
the computed value in the present formulation.  
The smallness of the coherence peak seen in the experiments is 
likely to be an effect of strong correlations in addition to disorder 
\cite{bang03,han03}. 

To conclude, it will be important to perform experiments to look for 
signals for a spin-charge decoupled state, a quantum protectorate and 
chiral d$_1$+id$_2$ superconductivity. 

\smallskip
The authors would like to thank Prof. K. Yoshimura and 
Prof. Y. Kobayashi for providing NMR/NQR data.




\end{multicols}


\begin{thebibliography}{99}
\vspace{-0.5cm}

\bibitem{takada03} K. Takada {\it et al.}, Nature {\bf 422}, 53 (2003).

\bibitem{terasaki97} I. Terasaki, Y Sasago and K. Uchinokura,
Phys. Rev. {\bf B 56}, R12685 (2003).

\bibitem{sakurai03} H. Sakurai, {\it et al.}, cond-mat/0304503.

\bibitem{cao03} G. Cao {\it et al.}, cond-mat/0305503. 

\bibitem{jin03} R. Jin {\it et al.}, cond-mat/0306066.

\bibitem{chou03} F. C. Chou {\it et al.}, cond-mat/0306659. 

\bibitem{schaak03} R. E. Schaak {\it et al.}, cond-mat/0305450.

\bibitem{yayu03} Y. Wang {\it et al.}, cond-mat/0305455.

\bibitem{kobayashi103} Y. Kobayashi, M. Yokoi and M. Sato,
cond-mat/0305649; cond-mat/0306264.

\bibitem{waki03} T. Waki {\it et al.},  cond-mat/0306036.

\bibitem{fujimoto03} T. Fujimoto {\it et al.},  cond-mat/0307127.

\bibitem{ishida03} K. Ishida {\it et al.},  cond-mat/0308506.

\bibitem{lynn03} J. W. Lynn {\it et al.}, cond-mat/0307263.

\bibitem{jorgensen03} J. D. Jorgensen {\it et al.}, cond-mat/0307627. 

\bibitem{ueland03} B. G. Ueland {\it et al.}, cond-mat/0307106.

\bibitem{lorenz03} B. Lorenz {\it et al.}, cond-mat/0308143. 

\bibitem{yang03} H. D. Yang  {\it et al.}, cond-mat/0308031.

\bibitem{baskaran103} G. Baskaran, Phys. Rev. Lett. {\bf 91}, 097003 (2003).

\bibitem{brijesh03} B. Kumar and B. S. Shastry,
cond-mat/0304210.

\bibitem{wang03} Q. H. Wang, D. H. Lee and P. A. Lee,
cond-mat/0304377.

\bibitem{ogata03} M. Ogata, cond-mat/0304405.

\bibitem{singh00} D. J. Singh, Phys. Rev. {\bf B 68}, 020503 (2003).

\bibitem{tanaka03} A. Tanaka and X. Hu, cond-mat/0304409.

\bibitem{ikeda03} H. Ikeda, Y. Nishikawa and K. Yamada,
cond-mat/0308472.

\bibitem{baskaran203} G. Baskaran, cond-mat/0306569. 

\bibitem{anderson00} P. W Anderson, Science {\bf 288}, 480 (2000). 

\bibitem{laughlin00} R. B. Laughlin and D. Pines, PNAS, {\bf 97}, 28 (2000).  

\bibitem{mackenzie03} A. P. Mackenzie and Y. Maeno, Rev. Mod. Phys.
{\bf 75} 657 (2003). 

\bibitem{abrikosov} A. A. Abrikosov and L. P. Gorkov, Sov. Phys. JETP, 
{\bf 12}, 1243 (1960). 

\bibitem{anderson01} P. W. Anderson, cond-mat/0108522. 
 
\bibitem{bang03} Y. Bang, M. J. Graf and A. V. Balatsky,
cond-mat/0307127.

\bibitem{han03} Qiang Han and Z. D. Wang,
cond-mat/0308160.  

\bibitem{mila89} F. Mila and T. M. Rice, Physica {\bf C 157}, 561 (1989).

\end{thebibliography}
\end{document}